\documentclass[12pt]{article}
\usepackage{amssymb}
\usepackage{epsfig,color}
\usepackage{pstricks,graphicx,epsfig,color,amssymb,amsmath,amscd}
\usepackage{cite}
\usepackage[]{graphicx}
\usepackage{makeidx}
\usepackage{amsmath}
\usepackage{nccmath}
\usepackage{setspace}

\usepackage[]{caption}  
\def\mn{{\mu\nu}}

\renewcommand\rho{\varrho}

\newcommand{\be}{\begin{eqnarray}}
\newcommand{\ee}{\end{eqnarray}}

\begin{document}
\begin{titlepage}
\title{
{Gravitational Baryogenesis: Problems and Possible Resolution}
{
}\\
\small{Presented at $6^{th}$ International Conference on Particle Physics and Astrophysics 
(ICCPA-2022)}
}
\author{
E. Arbuzova$^{a,b}$, A. Dolgov$^{b,c}$, K. Dutta$^{d}$,  R. Rangarajan$^{e}$
}

\maketitle
\begin{center}
$^a$Department of Higher Mathematics, Dubna State University,
Universitetskaya st.\,19, Dubna, Moscow region 141983, Russia;\\
$^b${Department of Physics, Novosibirsk State University,\\ 
Pirogova st.\,2, Novosibirsk, 630090 Russia}\\
$^c${{Bogolyubov Laboratory of Theoretical Physics, Joint Institute for Nuclear Research,\\
Joliot-Curie st.\,6, Dubna, Moscow region, 141980 Russia}}\\
 $^d$Department of Physical Sciences, Indian Institute of Science Education and Research Kolkata, 
 Mohanpur 741246, India; 
 \\
$^{e}$ Mathematical and Physical Sciences Division, School of Arts and Sciences, Ahmedabad University, Navrangpura, Ahmedabad 380009, India. 
\\
\end{center}

\begin{abstract}
The coupling of baryonic current to the derivative of the curvature scalar, $R$, inherent to gravitational baryogenesis (GBG), leads to 
 a fourth order differential equation of motion for $R$ instead of the algebraic one of General Relativity (GR). 
 The fourth-order differential equation is generically unstable. We consider a possible mechanism of stabilization of GBG by 
 modification of gravity, introducing an $R^2$-term into the canonical action of GR.
 It is shown that  this mechanism allows for stabilization of GBG with bosonic and fermionic baryon currents. 
  We have  established the region of the model parameters leading to stabilization of $R$. Still, the standard cosmology 
  would be noticeably modified. 
\end{abstract}

\thispagestyle{empty}
\end{titlepage}



\section{Introduction}
An excess of matter over antimatter in our Universe is crucial for our very existence and is well supported by various observations. 
The local Universe is clearly matter dominated. The amount of antimatter is very small and it can be explained as the result of high energy collisions in space.
On the other hand, matter and antimatter seem to have  similar properties, therefore we could expect a matter-antimatter symmetric universe.  
The existence of large regions of antimatter in our neighbourhood would produce high energy radiation 
{created by} matter-antimatter annihilation {on the boundaries between matter and antimatter domains}, 
which is not observed. A satisfactory model of our Universe should be able to explain the origin of the matter-antimatter asymmetry.
Any initial asymmetry at inflation could not solve the problem of observed excess of matter over antimatter, because the energy density associated with {the observed non-zero 
baryonic number density} would not allow for sufficiently long inflation. 


The term  baryogenesis is used to indicate the generation of {the excess of matter (baryons)  over antimatter 
(antibaryons) or vice versa}.

In 1967 Andrey Sakharov 
{formulated three conditions}
today know as Sakharov Principles \cite{Sakharov:1967dj}, {necessary}
to produce a matter-antimatter asymmetry {in the} 
initially symmetric universe. These conditions include: 
\begin{enumerate}
\item	Non-conservation of baryonic number; 
\item	Breaking of symmetry between particles and antiparticles;
\item	Deviation from thermal equilibrium.
\end{enumerate}
However, not all of three Sakharov Principles are strictly necessary. For example, spontaneous baryogenesis (SBG) and 
gravitational bayogenesis (GBG) do not demand an explicit C and CP violation and  can proceed in thermal equilibrium. Moreover, these mechanisms are
usually most efficient in thermal equilibrium.

The statement that the cosmological baryon asymmetry can be created by spontaneous baryogenesis in thermal equilibrium was mentioned in the 
original paper by A.~Cohen and D.~Kaplan in 1987 \cite{Cohen:1987vi} and in 
the subsequent papers by A.~Cohen, D.~Kaplan, and A.~Nelson 
\cite{Cohen:1988kt,Cohen:1991iu} (for a review see \cite{Dolgov:1991fr,Rubakov:1996vz,Riotto:1999yt,AD-30}). 

The term "spontaneous" is related to spontaneous breaking of underlying symmetry of the theory, 
which ensures the conservation of the total baryonic number in the unbroken phase. This symmetry is 
supposed to be spontaneously broken and in the broken phase the Lagrangian density  acquires the term
\be
{\cal L}_{SBG} =  (\partial_{\mu} \theta) J^{\mu}_B\, ,
 \label{L-SB}
 \ee
where $\theta$ is a (pseudo) Goldstone field, and $J^{\mu}_B$ is the baryonic current of matter fields, which becomes non-conserved
{as a result of the symmetry breaking}. 

For a spatially homogeneous field, $\theta = \theta (t)$, the Lagrangian is reduced to a simple form
\be
{\cal L}_{SBG} =   \dot \theta\, n_B\,, \     n_B\equiv J^0_B. 
\label{L-SB-hom}
\ee
Here ${n_B}$ is the baryonic number density, so 
it is tempting to identify ${\dot \theta}$ with the chemical potential, $ \mu_B$, of the corresponding system. 
 However, such an identification is questionable
 \cite{Arbuzova:2016qfh,Dasgupta:2018eha}. It depends upon the representation chosen for the fermionic fields
and is heavily based on the assumption ${\dot \theta \approx const}$. In 
Ref.~\cite{Arbuzova:2016qfh} 
the assumption ${\dot \theta \approx const}$ is relaxed. 

Stimulated by spontaneous baryogenesis  the idea of gravitational baryogenesis was put forward \cite{GBG-1}. The 
scenario of SBG was modified by the introduction of the coupling of the baryonic current to the derivative 
of the curvature scalar $ R$:
\be
{\cal S}_{GBG} =  - \frac{1}{M^2} \int d^4 x \sqrt{-g}\, (\partial_\mu  R ) J^\mu_B\, ,
\label{S-GBG}
\ee 
where $g$ is the determinant of the space-time metric tensor and 
the mass parameter $M$ determines the energy scale of baryogenesis. There are a lot of articles on the subject, and a partial list of references is included
in Refs. \cite{Lambiase:2006dq,Sadjadi:2007dx,Lambiase:2012tn,Fukushima:2016wyz,Odintsov:2016hgc}. According to these
 papers, the GBG mechanism can successfully explain the magnitude of the cosmological baryon asymmetry of the universe.  

However, it was argued in Refs.~\cite{Arbuzova:2016cem,Arbuzova:2017vdj}, that the back reaction of the created non-zero baryonic density 
{on the space-time curvature} leads to strong 
instability of the cosmological evolution. In this paper we show that the problem of stability can be solved by adding to the Hilbert-Einstein action 
the quadratic in curvature term generated by quantum corrections~\cite{Ginz-Kirzh-Lyub,Gurovich:1979xg}. 
The underlying gravitational action has the form:
\be
S_{Grav}= - \frac{M_{Pl}^2}{16\pi }  \int d^4 x\, \sqrt{-g} \left( R - \frac{R^2}{6M_R^2} \right),
\label{S-grav}
\ee
where $M_{Pl}=1.22\cdot 10^{19}$ GeV is the Planck mass, and we use
the metric signature $(+,-,-,-)$. As is known, the $R^2$-term leads to excitation of the scalar degree of freedom, named scalaron, and 
$M_R$ is the scalaron mass. In the very early universe the $R^2$-term can generate inflation~\cite{Starobinsky:1980te}, and density perturbations. The amplitude of the observed density 
perturbations demands that $M_R = 3 \cdot 10^{13}$ GeV~\cite{Gorbunov:2012ij}
{
if the scalaron is the inflaton.  Otherwise $M_R > 3 \cdot 10^{13}$ GeV {is allowed}.  
Below we presume that the scalaron is the inflaton.}

\section{{Instability problem of gravitational baryogenesis} \label{s-problems-GBG}} 

The essential ingredient of the spontaneous baryogenesis is the coupling of the baryonic current the derivative
of the curvature scalar $\partial_\mu R$ (\ref{S-GBG}). Taken over canonical cosmological
Friedmann-Lemaitre-Robertson-Walker background, this interaction can successfully  fulfil the task of
generating the proper value of the baryon asymmetry of the universe. 

However, any curvature dependent term in the Lagrangian of the theory would modify the equations of the 
General Relativity (GR). The modified GR equations have been analysed in Refs.~\cite{Arbuzova:2017vdj,Arbuzova:2016qfh}. Since interaction (\ref{S-GBG}) is not just linear in the
curvature term multiplied by a constant, it leads to higher order equations describing  evolution of
gravitational fields. Higher order equations of motion are typically unstable with respect to small perturbations. 
According to the results of Refs.~\cite{Arbuzova:2017vdj,Arbuzova:2016qfh}, it indeed happens in the frameworks
of the SBG scenario and the characteristic time of the exponential instability is much shorter than the cosmological time.
It creates serious problem for realisation of  the SBG mechanism.

In this work we suggest to consider possible stabilisation of SBG and have proved that it can be realised but
the resulting  cosmological model suffers from too large value of $R$, much larger than that in the classical
Friedmann cosmology. 
Possible ways to cure this shortcoming are mentioned.

\section{Stabilisation of gravitational baryogenesis in modified gravity \label{s-scalar}}

\subsection{Bosonic case. \label{ss-bosons}}

Let us first consider the case when  baryonic number is carried by a complex scalar field $\phi $~\cite{Arbuzova:2016cem}. The total action has the form:
\be \nonumber
S_{tot}[\phi] = - \int d^4 x\, \sqrt{-g} \left[ \frac{M_{Pl}^2}{16\pi } \left( R - \frac{R^2}{6M_R^2} \right) + \frac{1}{M^2} (\partial_{\mu} R) J^{\mu}_{(\phi)}  - 
g^{\mu \nu} \partial_{\mu}\phi\, \partial_{\nu}\phi^* + U(\phi, \phi^*)\right]  \\ 
+ S_{matt}\,,
\label{act-tot}
\ee
where $U(\phi, \phi^* )$ is the potential of field $\phi$ and $S_{matt}$ is the matter action which does not include the field $\phi$. 
 In Eq. \eqref{act-tot} $R(t)$ is the classical curvature field, while $\phi (\vec x, t)$ is the quantum operator of light scalar particles.

We assume that the potential $U(\phi,\phi^*)$ is not invariant with respect 
to phase transformation $\phi \rightarrow \exp{(i q\beta)} \phi $ and thus the corresponding current 
\be 
J^{ \mu}_{(\phi)} = i q\, g^{\mu \nu}(\phi^* \partial_{\nu}\phi - \phi \partial_{\nu}\phi^*)
\label{current}
\ee
is not conserved. Here $q$ is the baryonic number of field $\phi $. The non-conservation of the current is 
necessary for the proper performance of the model, otherwise $S_{GBG}$ in Eq. \eqref{S-GBG} can be integrated away by parts.

Varying action \eqref{act-tot} over $g^{\mu \nu}$ we come to the following equations:
\be \nonumber
&& \frac{M_{Pl}^2}{16\pi } \left[  R_{\mn} - \frac{1}{2}g_{\mn} R -
 \frac{1}{3M_R^2}\left(R_{\mn}-\frac{1}{4} g_{\mn}R+g_{\mn} D^2-  D_\mu D_\nu\right) R \right] \\ \nonumber
&&-\frac{1}{M^2}\left[ \left(R_{\mu \nu} - (D_{\mu}D_{\nu} - g_{\mu\nu}D^2)\right) D_{ \alpha} J^{\alpha}_{(\phi) } +
\frac{1}{2} g_{\mu\nu} J^{\alpha}_{(\phi)}\,D_{\alpha} R  - \frac{1}{2} \left(J_{(\phi) \nu} D_{\mu} R + J_{(\phi) \mu} D_{\nu} R\right)\right]  \\ \nonumber
&&- \frac{1}{2} \left(D_{\mu} \phi \, D_{\nu} \phi^* + D_{\nu} \phi \, D_{\mu} \phi^*\right) + 
\frac{1}{2} g_{\mu\nu} \left[ D_{\alpha} \phi \, D^{\alpha} \phi^* -  U(\phi )\right]
{{- (D_\mu\phi) (D_\nu \phi^*)}} \\ 
&&=  \frac{1}{2}\, T_{\mu\nu}^{(matt)}\, ,
\label{EoM}
\ee
where $D_\mu$ is the covariant derivative in metric $g_{\mu\nu}$ (of course, for scalars $D_\mu = \partial_\mu$) and
$T_{\mu\nu}^{(matt)}$ is the energy-momentum tensor of matter obtained from action $S_{matt}$.

Taking the trace of equation (\ref{EoM}) with respect to $\mu$ and $\nu$  and changing sign we obtain:
\be \nonumber
\frac{M_{Pl}^2}{16\pi }\, \left( R + \frac{1}{M_R^2} D^2 R \right)
+ \frac{1}{M^2}\left[ (R + 3 D^2) D_{\alpha} J^{ \alpha}_{(\phi)} + J^{\alpha}_{(\phi)} \,D_{\alpha}R  \right] - 
D_{\alpha} \phi \, D^{\alpha} \phi^* + 2 U(\phi ) \\ 
= - \frac{1}{2} \, T^{(matt)} { = 0} \,,
\label{trace-eq}
\ee
where $T^{(matt)} = g^{\mu \nu} T_{\mu\nu}^{(matt)}$ is the trace of the energy-momentum tensor of matter.
For the usual relativistic matter $T^{(matt)} = 0$, while for  scalar field $\phi $ the trace of the  energy-momentum tensor  is nonzero:
\be 
T_{\mu}^{\mu} (\phi ) = - 2D_{\alpha} \phi \, D^{\alpha} \phi^* + 4 U(\phi ) .
\label{trace-phi}
\ee
The equation of motion for field $\phi$ is:
\be
D^2 \phi + \frac{\partial U}{\partial \phi^*} = - \frac{i q}{M^2} \left(2 D_{\mu} R\, D^{\mu} \phi + \phi D^2 R\right)\, .
\label{EoM-phi}
\ee
According to definition (\ref{current}) and Eq. \eqref{EoM-phi}, the current divergence is:
\be 
D_{\mu} J^{\mu} =  \frac{2q^2}{M^2} \left[ D_{\mu} R\, (\phi^* D^{\mu}\phi + \phi D^{\mu}\phi^*) + |\phi|^2 D^2 R \right]
+ i q \left(\phi \frac{\partial U}{\partial \phi} - \phi^* \frac{\partial U}{\partial \phi^*} \right)\,.
\label{J-div}
\ee
For homogeneous curvature scalar $R(t)$ in spatially flat FLRW-metric 
\be
ds^2=dt^2 - a^2(t) d{\bf r}^2\, 
\label{ds-2}
\ee
Eq. \eqref{trace-eq} is reduced to: 
\be \nonumber
\frac{M_{Pl}^2}{16\pi }\, \left[ R + \frac{1}{M_R^2} (\partial_t^2 + 3 H \partial_t) R \right]
+ \frac{1}{M^2}\left[ (R + 3 \partial_t^2 + 9 H \partial_t) D_{\alpha} J^{\alpha}_{(\phi)} + 
\dot R \, J_{(\phi)}^0 \right]  \\
{{+ 2 U(\phi ) - (D_\alpha\phi) (D^\alpha \phi^*)  = 0.}}
\label{trace-eq-FRW}
\ee 
where  $J_{(\phi)}^0$ is the baryonic number density of the $\phi$-field,
$H = \dot a/a$ is the Hubble parameter, and the divergence of the current is given by the expression:
\be 
D_{\alpha} J^{\alpha}_{(\phi)} = \frac{2q^2}{M^2} \left[ \dot R\, (\phi^* \dot \phi + \phi \dot \phi^*) + 
(\ddot R + 3H \dot R)\, \phi^*\phi\right]
+ i q \left(\phi \frac{\partial U}{\partial \phi} - \phi^* \frac{\partial U}{\partial \phi^*} \right)\,.
\label{J-div-hom}
\ee
 As we see in what follows, the last two terms in Eq. \eqref{trace-eq-FRW} do not have an essential impact on the cosmological instability 
found in Ref.~\cite{Arbuzova:2016cem} and will be disregarded below.

Let us note that the statement of exponential instability of $R(t)$ \cite{Arbuzova:2016cem} does not depend on the conservation or non-conservation of the 
current
from the potential term $\left(\phi{\partial U}/{\partial \phi} - \phi^* {\partial U}/{\partial \phi^*}\right) $
in Eq.~\eqref{J-div-hom}.
However if 
the current from
this term 
is conserved then the baryon asymmetry is not generated. 
On the other hand the 
term in square brackets in Eq.~\eqref{J-div-hom} does not lead to generation of the baryon asymmetry but leads to exponential instability of $R(t)$. 
Below we ignore the last term of Eq.~\eqref{J-div-hom}.

Performing thermal averaging of the 
normal ordered
bilinear products of field $\phi $  
in the high temperature limit
(see Appendix of Ref.~\cite{Arbuzova:2016cem})
 in accordance with equations: 
 \be
\langle \phi^* \phi \rangle =  \frac{T^2}{12}\, , \ \ \ 
\langle \phi^* \dot \phi + \dot \phi^* \phi \rangle = 0\, ,
\label{av-phi}
\ee
{
and using Eq. \eqref{J-div-hom} 
}
we obtain the fourth order differential equation:
\be
\frac{M_{Pl}^2}{16\pi }\, \left( R + \frac{1}{M_R^2} D^2 R \right) + 
\frac{q^2}{6 M^4} \left(R + 3 \partial_t^2 + 9 H \partial_t \right)
\left[ 
\left(\ddot R + 3H  \dot R\right) T^2 \right] + 
\frac{1}{M^2} \dot R \, \langle J_{(\phi)}^0 \rangle \nonumber\\
=
- 2 U(\phi ) + (D_\alpha\phi) (D^\alpha \phi^*).
\label{trace-eq-plasma}
\ee

Here $\langle J_{(\phi)}^0 \rangle $ is the thermal average value of the baryonic number density of $\phi$, which is supposed to vanish initially, but created 
through the process of the gravitational baryogenesis. This term can be neglected because the {baryon}
asymmetry is normally quite small. 
Even if it is not small it does not have considerable impact on the explosive rise of the curvature scalar. As we see in what follows the evolution of $R(t)$
proceeds much faster than the cosmological evolution, that is $\ddot R / \dot R \gg H$. 
Consequently, we neglect the terms proportional to $R$ with respect to the terms proportional to the second derivative of $R$, $\ddot R$. 
We also consider the terms of the type $HR$ as small w.r.t. to $dR/dt$.
We can check that this presumption is true a posteriori 
with the obtained solution for $R(t)$.  

Keeping only the dominant terms we simplify the above equation to: 
\be
\frac{d^4 R}{dt^4} + \frac{\kappa^4}{M_R^2}\frac{d^2 R}{dt^2}  + \kappa^4 R =  - 
{T_{\mu}^{\mu} (\phi )}
\frac{M^4}{q^2 T^2}
,  
\label{d4-R}
\ee
where 
\be 
\kappa^4 = \frac{M_{Pl}^2 M^4}{8 \pi q^2 T^2}\,.
\label{mu}
\ee
While studying the instability of the solution 
we do not take into account the r.h.s. of Eq. (17) which does not depend upon R.
Looking for the solution of Eq.~\eqref{d4-R}   in the form 
$R = R_{in} \exp (\lambda t)$, 
we obtain the characteristic equation: 
 \be
 \lambda^4 + \frac{\kappa^4}{M_R^2} \lambda ^2 + \kappa^4 =0 
 \label{lambda-4}
 \ee
 with 
 {the eigenvalues}
 {$\lambda$}
 defined by the expression:
 \be 
 \lambda ^2 = - \frac{\kappa^4}{2M_R^2} \pm \kappa^2 \sqrt{\frac{\kappa^4}{4M_R^4} - 1}.
 \label{lambda-2}
 \ee

There is no instability, if $ \lambda ^2 < 0$ and Eq. \eqref{d4-R} has only oscillating solutions. It is realised, if $\kappa^4 >  4M_R^4$.  
Using the expression in Eq.~\eqref{mu} for $\kappa^4$ and taking $M_R = 3 \cdot 10^{13}$ GeV we find the stability condition:
\be
M > 3 \cdot 10^4 \,\text{GeV} \left(\frac{q\,T}{\text{GeV}}\right)^{1/2},
\label{stab-cond}
\ee
 which is fulfilled for all interesting values of $M$. 

The value of $\lambda $ depends upon the relation between $\kappa $ and $M_R$. If $\kappa \sim M_R$ then the frequency of the oscillations
 of curvature  is of the order of $M_R$ {and}
 $|\lambda|\sim M_R$. If $\kappa \gg M_R$ then there are two possible solutions $|\lambda|\sim M_R$ and 
 {
 $|\lambda|\sim \kappa (\kappa /M_R) \gg M_R$. }
 High frequency oscillations of $R$ would lead to efficient gravitational particle production and, as a result, to 
 damping of the oscillations.

\subsection{Fermionic case}

In this section we consider the case when baryonic number is carried by fermions. 
The gravitational part of the action has the form as in Eq.~\eqref{S-grav}, while the fermionic part of the action is 
{the same} as in Refs. \cite{Arbuzova:2017vdj,Dasgupta:2018eha}:
 \begin{eqnarray} \nonumber
{{ {\cal L} [Q, L]}} &= &\frac{i}{2} (\bar Q \gamma ^\mu \nabla _\mu Q -   \nabla _\mu \bar Q\, \gamma ^\mu Q) - m_Q\bar Q\,Q\\ \nonumber
&+& 
\frac{i}{2} (\bar L \gamma ^\mu \nabla _\mu L -   \nabla _\mu \bar L \gamma ^\mu L)
- m_L\bar L\,L \\ 
&+& \frac{g}{m_X^2}\left[(\bar Q\,Q^c)(\bar Q L) + (\bar Q^cQ)(\bar L Q) \right]
+ \frac{d }{M^2} (\partial_{\mu} R) J^{\mu} +{ {\cal L}_{matt}}\,, 
\label{L-matt}
\end{eqnarray}
where $Q$ is the  quark-like field with non-zero baryonic {number $B_Q$, $Q^c$} 
is the charged conjugated quark operator,  $L$ is
another fermionic field (lepton),, and
$\nabla_\mu  $ is the covariant derivative of the Dirac fermions in tetrad formalism.
The quark current is  $J^{\mu } = B_Q\bar Q \gamma ^{\mu} Q$  with $\gamma ^{\mu}$ being the curved space gamma-matrices,  and
${ {\cal L}_{matt}}$ describes all other forms of matter. The four-fermion interaction between quarks and leptons is introduced to ensure the
necessary non-conservation of the baryon number with $m_X$ being a constant
parameter with dimension of mass and $g$ being a dimensionless coupling constant. In the term, describing interaction of the baryonic current of  
fermions with the derivative of the curvature scalar, $M$ is a constant parameter with dimension of mass 
and $d=\pm 1$ is dimensionless coupling constant which is introduced to allow for an arbitrary sign of the above expression.

Gravitational equations of motion with an account of $R^2/M_R^2$-term in analogy with Eq.~\eqref{EoM} take the form: 
\be \nonumber
&&\frac{M_{Pl}^2}{8\pi }\left[ R_{\mu\nu} - \frac{1}{2} g_{\mu\nu} R 
- \frac{1}{3M_R^2}\left(R_{\mn}-\frac{1}{4} g_{\mn}R+g_{\mn} D^2-  D_\mu D_\nu\right) R
\right] \\ \nonumber
&=& \frac{g_{\mu \nu}}{2} \frac{g}{m_X^2}\left[(\bar Q\,Q^c)(\bar Q L) + (\bar Q^cQ)(\bar L Q) \right] \\ \nonumber
 &+&
\frac{i}{4} \left[\bar Q (\gamma _\mu \nabla _\nu + \gamma _\nu \nabla _\mu ) Q
-  (\nabla _\nu \bar Q\, \gamma _\mu  + \nabla _\mu \bar Q\, \gamma _\nu )Q  \right] \\ \nonumber
& + &
\frac{i}{4} \left[\bar L (\gamma _\mu \nabla _\nu + \gamma _\nu \nabla _\mu ) L
 -    (\nabla _\nu \bar L\, \gamma _\mu  + \nabla _\mu \bar L\, \gamma _\nu )L  \right]  \\ 
 & - &
 \frac{2d}{M^2} \left[R_{\mu\nu} + g_{\mu \nu } D^2 - D_{\mu} D_{\nu}\right] D_{\alpha} J^{\alpha} + 
 \frac{d}{2M^2} \left(J_{\mu} \partial_{\nu}R +  J_{\nu} \partial_{\mu}R \right) + T_{\mu\nu}^{matt}\,.
 \label{EoM-grav}
\ee
Taking the trace of Eq. (\ref{EoM-grav}) with an account of fermion equations of motion we obtain:
\begin{eqnarray} \nonumber
- \frac{M_{Pl}^2}{8\pi }\, \left( R + \frac{1}{M_R^2} D^2 R \right) = m_{Q} \bar Q Q + m_L \bar L L 
+  \frac{2g}{m_X^2}\left[(\bar Q\,Q^c)(\bar Q L) + (\bar Q^cQ)(\bar L Q) \right] \\
 -  \frac{2d}{M^2} (R + 3D^2) D_{\alpha} J^{\alpha} + { T_{matt}}\,,
 \label{trace}
 \end{eqnarray}
 where $ {T_{matt}}$ is the trace of the energy momentum tensor of all other fields.  In the early 
 {universe when} various species are relativistic,
 we can take  $T_{matt} = 0$. The average  expectation value of the quark-lepton interaction term proportional to $g$ is also 
small, so the contribution of all matter fields may be neglected   and hence the only  term which remains in the r.h.s. 
of Eq. \eqref{trace} is that proportional to $D_{\alpha} J^{\alpha}$.

A higher order differential equation
 for $R$ {is obtained}
after we substitute the current divergence, $D_{\alpha} J^{\alpha}$, calculated 
from the kinetic equation in the external field $R$ \cite{Arbuzova:2017vdj}, { into Eq. \eqref{trace}.}
For the spatially homogeneous case 
\begin{eqnarray}
D_{\alpha} J^{\alpha} = (\partial_t +3 H) n_B = I_B^{coll}, 
\label{kin-eq-gnrl}
\end{eqnarray}
where the collision integral, $I_B^{coll}$,  in the lowest order of perturbation theory is equal to:
\begin{eqnarray} 
&&I^{coll}_B =- 3 B_q (2\pi)^4  \int  \,d\nu_{q_1,q_2}  \,d\nu_{\bar q_3, l_4}
\delta^4 (q_1 +q_2 -q_3 - l_4)
\nonumber\\
&& \left[ |A( q_1+q_2\rightarrow  \bar q_3 +l_4)|^2
f_{q_1} f_{q_2} -
 |A( \bar q_3 +l_4  \rightarrow  q_1+q_2 ) |^2
 f_{\bar q_3} f_{l_4}
 \right].
\label{I-coll}
\end{eqnarray}
Here $ A( a \rightarrow b)$ is the amplitude of the transition from state $a$ to state $b$,
$B_Q$ is the baryonic number of quark, $f_a$ is the phase space distribution (the
occupation number), and 
\begin{eqnarray}
d\nu_{q_1,q_2}  =  
\frac{d^3 q_1}{2E_{q_1} (2\pi )^3 }\,  \frac{d^3 q_2}{2E_{q_2} (2\pi )^3 } ,
\label{dnuy}
\end{eqnarray}
where $E_q = \sqrt{ q^2 + m^2}$ is the energy of particle with three-momentum
$q$ and mass $m$. The element of phase space of final particles, $d\nu_{\bar q_3, l_4} $, is defined analogously. 

We choose such representation of the quark operator, $Q$, for which the interaction  of baryonic current with the derivative of the 
curvature scalar in Eq. \eqref{L-matt} vanishes but reappears in the  quark-lepton interaction term: 
\begin{eqnarray}
 \frac{2g}{m_X^2}\left[ e^{-3id B_Q R/M^2}\, (\bar Q\,Q^c)(\bar Q L) + 
 e^{3id B_QR/M^2}\,(\bar Q^cQ)(\bar L Q) \right] .
\label{int-R}
\end{eqnarray}
We make the simplifying assumption that the evolution of $R$ can be approximately described by the law
\be
R(t) \approx R(t_0) +  (t - t_0) \dot R.
\label{R-of-t} 
\ee
We assume that $\dot R(t)$ slowly changes at the characteristic time scale of the reactions, which contribute to the collision integral 
\eqref{I-coll}, and so we can approximately take   $\dot R \approx const$. 

According to the rules of quantum field theory the reaction probability is given by the square of the integral over space and time of 
the amplitude of the corresponding process. In the case of time independent interaction it leads to the energy conservation, 
$\Sigma E_{in} =  \Sigma E_{fin}$. If the interaction depends upon time the energy evidently is non-conserved and in our case,
e.g. for the reaction $q_1+q_2\rightarrow  \bar q_3 +l_4$,  the energy balance has the form:
\be 
E(q_1) + E(q_2) = E(q_3) + E(l_4) + 3d B_Q\dot R /M^2.
\label{balance}
\ee

In kinetic  equilibrium the phase space distribution of fermions has the form 
\begin{eqnarray}
f = \frac{1}{e^{(E/T - \xi)} + 1} \approx e^{-E/T + \xi}, 
\label{f-eq}
\end{eqnarray}
where $\xi = \mu/T$ is the dimensionless chemical potential, different for quarks, $\xi_q$,
and leptons, $\xi_l$. In thermal equilibrium case the condition of conservation of chemical potentials is fulfilled, that is 
$\Sigma\, \xi_{in} = \Sigma\, \xi_{fin}$. In particular it demands that chemical potentials of particles and antiparticles are 
equal by magnitude and have opposite signs: $\xi = - \bar \xi$, as follows e.g. from the consideration of particle-antiparticle 
annihilation into 
{ different numbers of}
photons.  If energy is not conserved, {due to time-dependent $R(t)$,} 
the conservation of chemical potentials is also 
broken, as we see in what follows. 

We assume that $\xi \ll 1$ and hence distribution \eqref{f-eq} turns into:
\be
f \approx e^{-E/T}(1+\xi).
\label{ksi-small}
\ee
We also assume that $3d\, B_Q\dot R /(M^2\,T) \ll 1$ and correspondingly the balance of chemical potentials in equilibrium 
 for the reactions $q_1+q_2\leftrightarrow  \bar q_3 +l_4$ leads to:
  {
 \be
 3 \xi_q - \xi_l - \frac{3d\, B_Q\dot R(t)}{M^2\,T} = 0.
 \label{ksi-balance} 
 \ee
 }
 
{
Following Ref.~\cite{Arbuzova:2017vdj}, we express
\be
n_B \approx \frac{g_s B_Q}{6}\xi_q T^3,
\ee
where $g_s$ is the number of quark spin states.  Since we are
studying instability of $R$ whose timescale is presumed to be much smaller
than the expansion rate of the Universe, we approximate
\be
D_\alpha J^\alpha \approx \dot n_B \approx \frac{g_s B_Q}{6}\dot\xi_q T^3
\\
\approx \frac{g_s B_Q}{6}\dot\xi_q^{eq} T^3,
\label{DalphaJalpha}
\ee
$\xi_q^{eq}$ is obtained from Eq.~\eqref{ksi-balance}, using the
conservation of
the sum of baryonic and leptonic numbers which implies $\xi_l = -\xi_q/3 $.
Then
\be
\xi_q^{eq}= \frac{9d\, B_Q\dot R(t)}{10 M^2\,T}\,.
\label{xi-eq}
\ee
Substituting Eq.~\eqref{xi-eq} in Eq.~\eqref{DalphaJalpha} and neglecting the
$\dot T$-term, Eq.~\eqref{trace} gives
the following fourth order differential 
equation for the curvature scalar:
}
\be
\frac{d^4 R}{dt^4} + \frac{\kappa_f^4}{M_R^2}\frac{d^2 R}{dt^2}  + \kappa_f^4 R =   0 , 
\label{d4-R-f}
\ee
where 
\be
\kappa_f^4 = \frac{5 M_{Pl}^2 M^4 }{36\pi  g_s B_Q^2 T^2}\,. 
\label{kappa-f}
\ee
Once again, we consider terms containing $R$ as small with respect to the terms containing $\ddot R$. 
The value of $\kappa_f$ is only slightly numerically different from $\kappa$ in 
Eq.~\eqref{mu} and has the same dependence upon the essential parameters, so 
the solutions of Eqs. \eqref{d4-R} and \eqref{d4-R-f} practically coincide.

\section{{Discussion}}

We have shown that discovered in Refs.~\cite{Arbuzova:2016cem, Arbuzova:2017vdj}
exponential instability of the curvature scalar inherent to the mechanism of spontaneous baryogenesis can be
successfully cured in modified gravity. The special form of gravity modification {by introduction} of $R^2$-term
into canonical Hilbert-Einstein action of General Relativity was  explored as a workable mechanism. 

However, the stabilized asymptotic value of $R$ is extremely large and together with possibly successful baryogenesis
would still strongly perturb canonical cosmology. Possible ways out of this problem 
could either be a more
complicated model of $F(R)$ gravity or a proper account of particle production created by high frequency oscillations
of $R(t)$. Both options open interesting possibilities for
future research.

\end{document}